\pdfoutput=1 
\documentclass[11pt,a4paper]{article}
\usepackage{graphicx}
\usepackage{caption}
\usepackage{subcaption}
\usepackage{jinstpub}
\usepackage[separate-uncertainty,retain-explicit-plus,per-mode=symbol,binary-units]{siunitx}
\usepackage{array,mathtools,dcolumn}
\usepackage{rotating}
\usepackage[version=4]{mhchem}
\DeclareSIUnit\year{y}
\DeclareSIUnit\inch{in}
\DeclareSIUnit\c{\mbox{$c$}}
\DeclareSIUnit\electron{\mbox{$e^-$}}

\newcommand{\mbb}{\mbox{$m_{\beta\beta}$}}

\newcommand{\bb}{\mbox{$\beta$$\beta$}}

\title{Measurement of the ionization response of amorphous selenium with 122\,keV $\gamma$ rays}
\author[a]{X.~Li,}
\author[b]{A.E.~Chavarria,}
\author[c]{S.~Bogdanovich,}
\author[a]{C.~Galbiati,}
\author[b]{A.~Piers}
\author[c]{and B.~Polischuk}

\affiliation[a]{Department of Physics, Princeton University, Princeton, New Jersey, United States}
\affiliation[b]{Center for Experimental Nuclear Physics and Astrophysics, University of Washington, Seattle, Washington, United States}
\affiliation[c]{Detector Technologies, Hologic Corporation, Newark, Delaware, United States}

\emailAdd{xinranli@princeton.edu}
\abstract{
We performed a measurement of the ionization response of \SI{200}{\micro\meter}-thick amorphous selenium (aSe) layers under drift electric fields of up to \SI{50}{\volt\per\micro\meter}.
The aSe target was exposed to ionizing radiation from a \ce{^{57}Co} radioactive source and the ionization pulses were recorded with high resolution.
Using the spectral line from the photoabsorption of \SI{122}{\kilo\electronvolt} $\gamma$ rays, we measured the charge yield in aSe and the line width as a function of drift electric field.
From a detailed microphysics simulation of charge generation and recombination in aSe, we conclude that the strong dependence of recombination on the ionization track density provides the dominant contribution to the energy resolution in aSe.
These results provide valuable input to estimate the sensitivity of a proposed next-generation search for the neutrinoless $\beta\beta$ decay of \ce{^{82}Se} that aims to employ imaging sensors with an active layer of aSe.
We estimate the RMS line width of the integrated ionization signal from neutrinoless \bb\ decay events (of deposited energy \SI{3.0}{\mega\electronvolt}) to be 2.0\% for a drift field of \SI{50}{\volt\per\micro\meter}.
The energy resolution can be improved to 1\% by correcting for the charge yield as a function of ionization density along the imaged electron tracks.
}

\begin{document}
\maketitle
\flushbottom

\section{Introduction}
\label{sec:intro}

Imaging sensors made from an ionization target layer of amorphous selenium (aSe) coupled to a silicon complementary metal-oxide-semiconductor (CMOS) active pixel array for charge readout were proposed as a promising technology to search for the neutrinoless \bb\ decay of \ce{^{82}Se}~\cite{Chavarria:2016vw}.
The preliminary study in Ref.~\cite{Chavarria:2016vw} suggests that a detector consisting of a large array of these devices could achieve background levels smaller than \SI{1E-6}{\per\kilo\gram\per\year}, which would allow to probe Majorana neutrino masses (\mbb ) as low as \SI{1}{\meV\per\square\c} for an almost definitive test on the origin of the neutrino mass~\cite{DellOro:2016gf}.
Ref.~\cite{Chavarria:2016vw} made a series of assumptions on the expected performance of the devices from the available literature in medical imaging, where aSe x-ray detectors have been traditionally employed for mammography.
Under those assumptions, the proposed technology combines the precise energy resolution required to reject background from the two-neutrino \bb\ decay channel, and the efficient determination of the event topology necessary for a powerful rejection of $\alpha$, $\beta$ and $\gamma$-ray backgrounds from natural radioactivity.
A significant uncertainty in the proposal is the intrinsic ionization response of aSe, which determines the fundamental limit on the energy resolution at the \bb\ decay endpoint ($Q_{\beta\beta}=3.0$\,\si{\mega\electronvolt}).
In this paper, we present results on the ionization response of aSe to the \SI{122}{\kilo\electronvolt} $\gamma$ rays emitted by a \ce{^{57}Co} radioactive source.
These measurements provide crucial input for the understanding of the charge generation and transport properties of aSe, which will allow for a more realistic model of the performance of the proposed detector to search for neutrinoless \bb\ decay.

\section{Experimental setup}
\label{sec:setup}

\begin{figure}[t!]
\centering
\includegraphics[width=\textwidth]{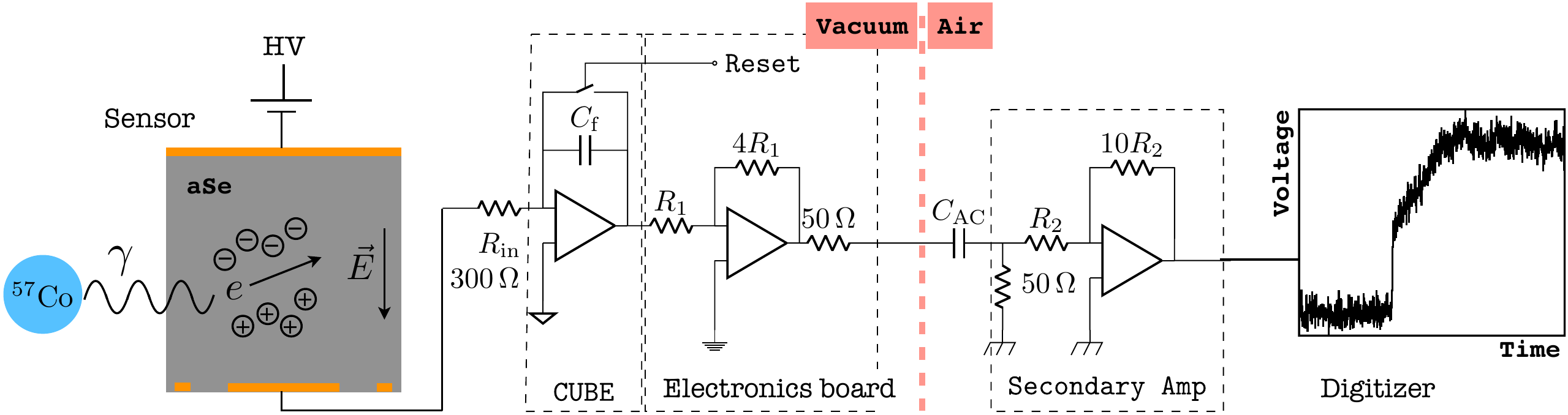}
\caption{Schematic of the experimental setup. A photoelectron $e$ from the absorption of $\gamma$ rays from a \ce{$^{57}$Co} source generates free e-h pairs in a target layer of aSe. The free charge is drifted by an electric field $\vec{E}$ and collected by electrodes. The cathode is connected to an electronics chain whose output voltage signal is proportional to the charge collected from the ionization event, which is digitized and recorded for analysis.
The labeled electrical elements are the input resistor ($R_{in}$), the CUBE feedback capacitance ($C_f$), the gain resistors of the amplification stages ($R_1$ and $R_2$), and the signal coupling capacitor ($C_{AC}$).
}
\label{fig:fullschematic}
\end{figure}

We built a single-pixel detector to read out the ionization signal generated by $\gamma$ rays incident on an aSe target layer.
The detector consisted of an aSe layer sandwiched between two gold electrodes.
A high voltage (HV) was applied to the anode to create an electric field across the aSe layer toward the cathode, which was connected to the input of a CMOS preamplifier whose potential was close to ground.
Electron-hole (e-h) pairs produced by ionizing radiation in the aSe were drifted toward and collected by the electrodes.
An electronics chain, starting with the CMOS preamplifier, resulted in a voltage signal proportional to the integrated current out of the cathode.
The output voltage was digitally sampled to record time traces of the ionization signal.
Figure~\ref{fig:fullschematic} shows a sketch of the experimental setup.

\begin{figure}[t!]
\centering
\includegraphics[width=0.85\textwidth]{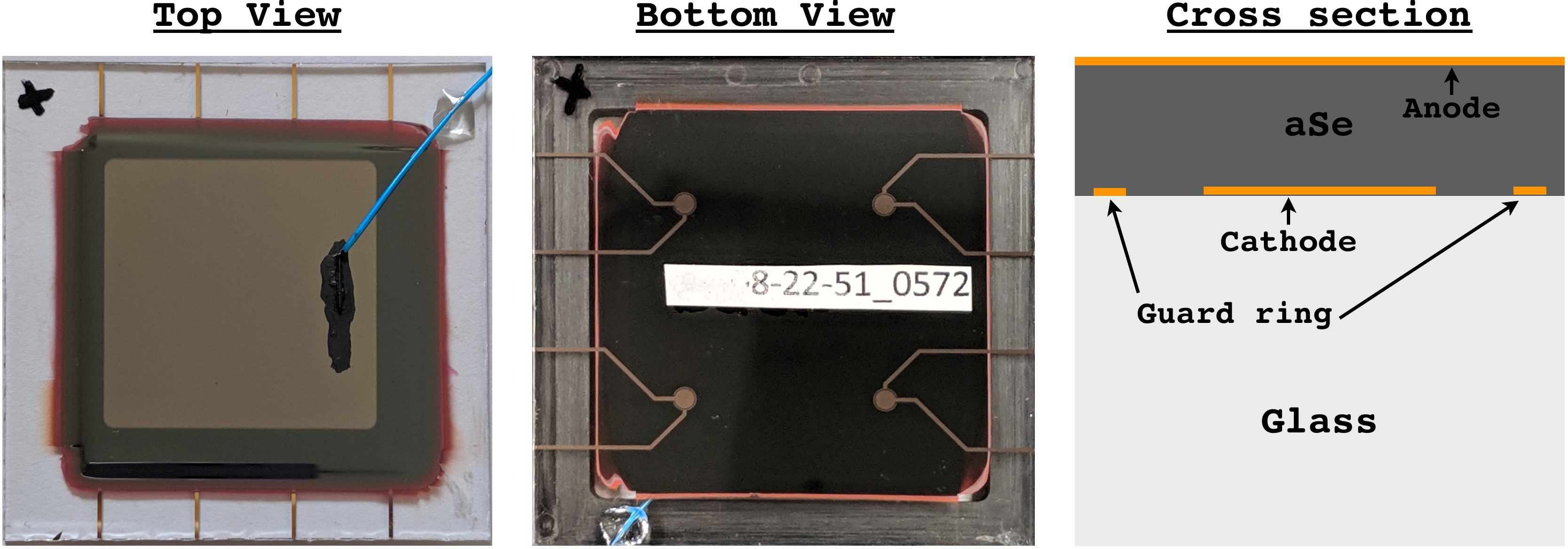}
\caption{Photographs of the aSe device with four sensors, and a cross section of one of the sensors. The black region with dark red edges seen from the top is the aSe layer. The top anode is very thin and appears as a light shaded square on top of the aSe. The blue HV cable is attached to the anode with carbon paint. The gold cathodes and guard rings with connecting traces can be seen from the bottom through the glass slide.}
\label{fig:sensors}
\end{figure}

Device fabrication started with four \SI{2}{\milli\meter}-diameter gold cathodes surrounded by concentric guard rings deposited on a glass slide.
The glass slide was then sent to Hologic Corporation for the deposition of a \SI{200}{\micro\meter}-thick aSe layer on top of the cathodes.
A final gold deposition covered the entire top surface of the aSe to act as the anode.
A hole blocking layer is present between the anode and the aSe to prevent hole injection and minimize leakage current.
Thin gold traces on the glass slide connected to the cathodes and guard rings extend beyond the aSe to the edge of the glass slide for electrical connection.
The total capacitance of the individual sensors is \SI{<3}{\pico\farad} to maximize signal-to-noise.
Figure~\ref{fig:sensors} shows one of the fabricated devices with the four sensors.

The cathode of the sensor was connected to the CUBE pulse-reset charge sensitive preamplifier with \SI{50}{\femto\farad} feedback capacitance.
The CUBE CMOS chip was glued and wire bonded to a carrier board that was mounted on a plastic holder together with the aSe device.
The connection between one of the cathode traces of the aSe device to the input trace of the CUBE carrier board was done by pressing a pin connector soldered on the carrier board firmly onto the trace on the edge of the glass slide.
The guard ring was likewise connected to the ground of the carrier board.
This method allowed to easily test different aSe sensors while minimizing the input capacitance.
The HV connection to the device was done with a single wire fixed on the anode with carbon paint.

\begin{figure}[t!]
\centering
\includegraphics[width=0.6\textwidth]{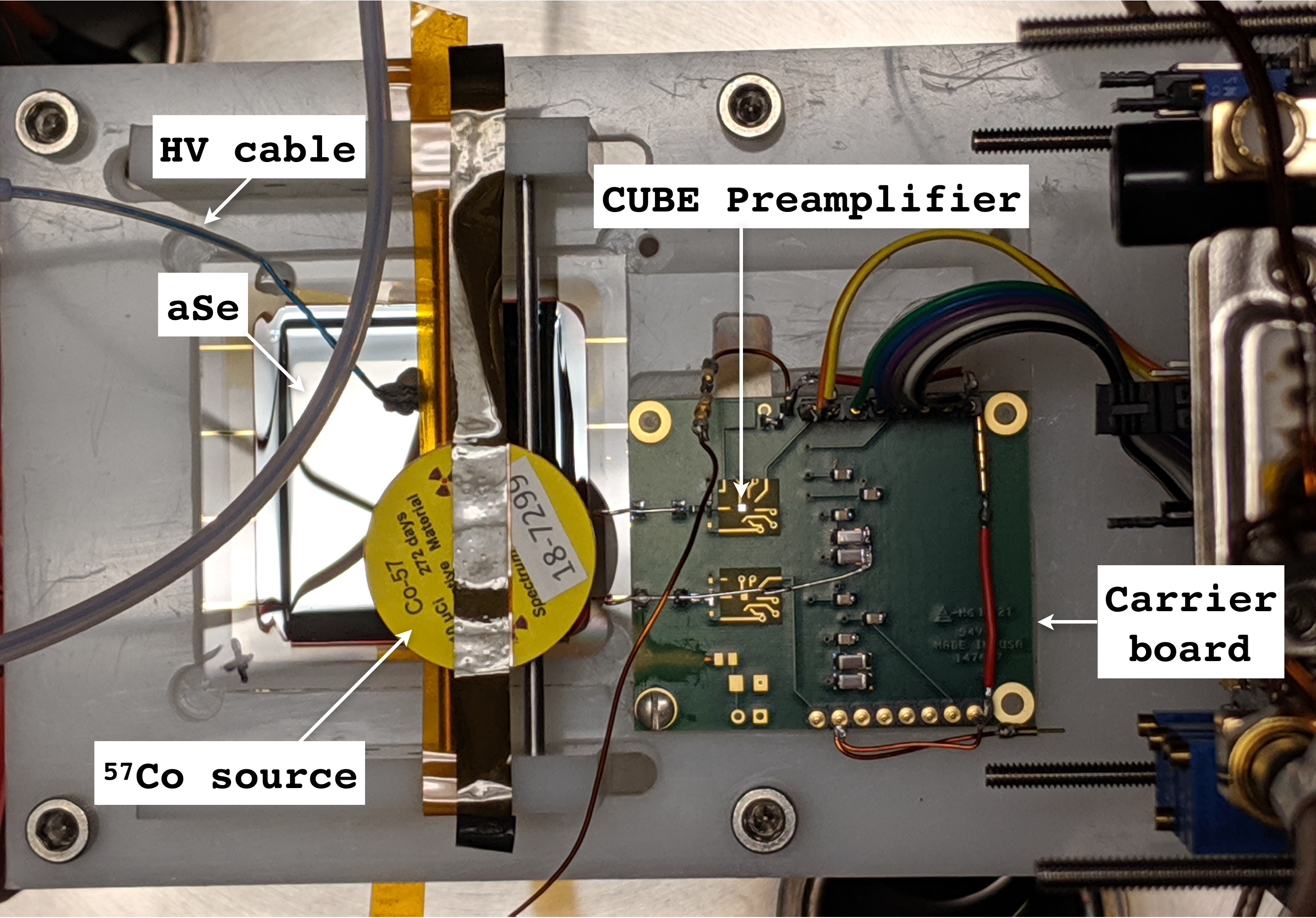}
\caption{Picture of the components mounted on the plastic holder inside the vacuum chamber: aSe device, \ce{$^{57}$Co} source, collimator (hidden behind the source), and the carrier board with the CUBE preamplifier. The carrier board can accommodate two channels but only one is used for this experiment.}
\label{fig:setup}
\end{figure}

A \SI{1.3}{\milli\meter}-thick brass collimator with a \SI{1.0}{\milli\meter}-diameter hole was mounted above the connected sensor on the device.
The collimator was aligned with the sensor by placing a light source below the aSe device and aligning the shadow of the cathode with the hole of the collimator.
A \ce{$^{57}$Co} button source of estimated activity \SI{3}{\micro Ci} at the time of measurement was then taped on top of the collimator.
Figure~\ref{fig:setup} shows the plastic holder with all mounted components.

The carrier board of the CUBE was connected with a ribbon cable to an electronics board that amplified the signal, and provided biases and reset pulses to the CUBE.
To minimize noise, the CUBE does not have a feedback resistor and its feedback capacitor is instead discharged by a reset pulse generated by the CUBE electronics every time a saturation threshold is reached.
Thus, the CUBE acted as an integrator where the output signal was proportional to the charge collected over time, and the frequency of the reset pulses was proportional to the leakage current of the sensor.

The plastic holder with the aSe device, \ce{$^{57}$Co} source, and the CUBE carrier board, as well as the CUBE electronics board, were housed in a vacuum chamber, which was pumped down below \SI{1e-4}{\milli\bar} so that voltages up to \SI{10}{\kilo\volt} could be applied across the aSe layer without discharges.
The vacuum chamber also significantly reduced noise by acting as a Faraday cage to isolate the system from electromagnetic waves, and by stopping sound waves and vibrations that may produce microphonic noise.

A vacuum feedthrough was used to connect to the outside power supply of the electronics board, the HV, and to bring out the signal.
The HV was provided by a Stanford Research PS365 power supply.
The output signal was capacitively coupled with a high pass filter with $\tau\sim$~\SI{100}{\micro\second} time constant to remove the linearly increasing baseline of the CUBE output caused by leakage current.
The signal was then fed into a secondary linear amplifier to match the \SI{2}{\volt} dynamic range of the 12-bit CAEN V740 digitizer, which sampled the signal in digitizer units (ADU) at \SI{62.5}{\mega\hertz}.
The calibration constant of the sampled data was calculated from the specifications of the components in the electronics chain to be $G=7.9$~$e^-$/ADU. The digitizer was triggered by a dedicated module whose input was the signal from the secondary amplifier duplicated by a linear fan-in/fan-out unit.
The signal was fed to a shaping amplifier and a trigger was generated by a discriminator when the shaping amplifier output was 5 times above the RMS of the baseline level.
Because the CUBE reset pulses saturate the secondary amplifier, a gate was opened in response to the CUBE reset to inhibit the trigger for \SI{1}{\milli\second}.

Data runs were acquired with HV from \SIrange{2}{10}{\kilo\volt} in \SI{1}{\kilo\volt} steps.
After the HV was increased on the sensor, there was a high leakage current transient that decayed away after several hours.
The asymptotic leakage current level was \SI{<1}{\pico\ampere} at \SI{2}{\kilo\volt}, increasing with HV to \SI{15}{\pico\ampere} at \SI{8}{\kilo\volt}.
Once the sensor was stable, data was acquired continuously for several hours.
The shaping time of the amplifier in the trigger module was adjusted for each HV with the aid of an oscilloscope to match the peak time of the signal pulses, which changes due to the different drift time of the free charge in the aSe. 
Once triggered, the digitizer acquired a time trace in a time window from \SIrange{-10}{30}{\us} relative to the trigger time.

\section{Ionization signal pulses}
\label{sec:signal}

\begin{figure}
        \centering
        \begin{subfigure}[b]{0.75\linewidth}
        \includegraphics[width =\linewidth, trim={0 0 0 0.9cm}, clip]{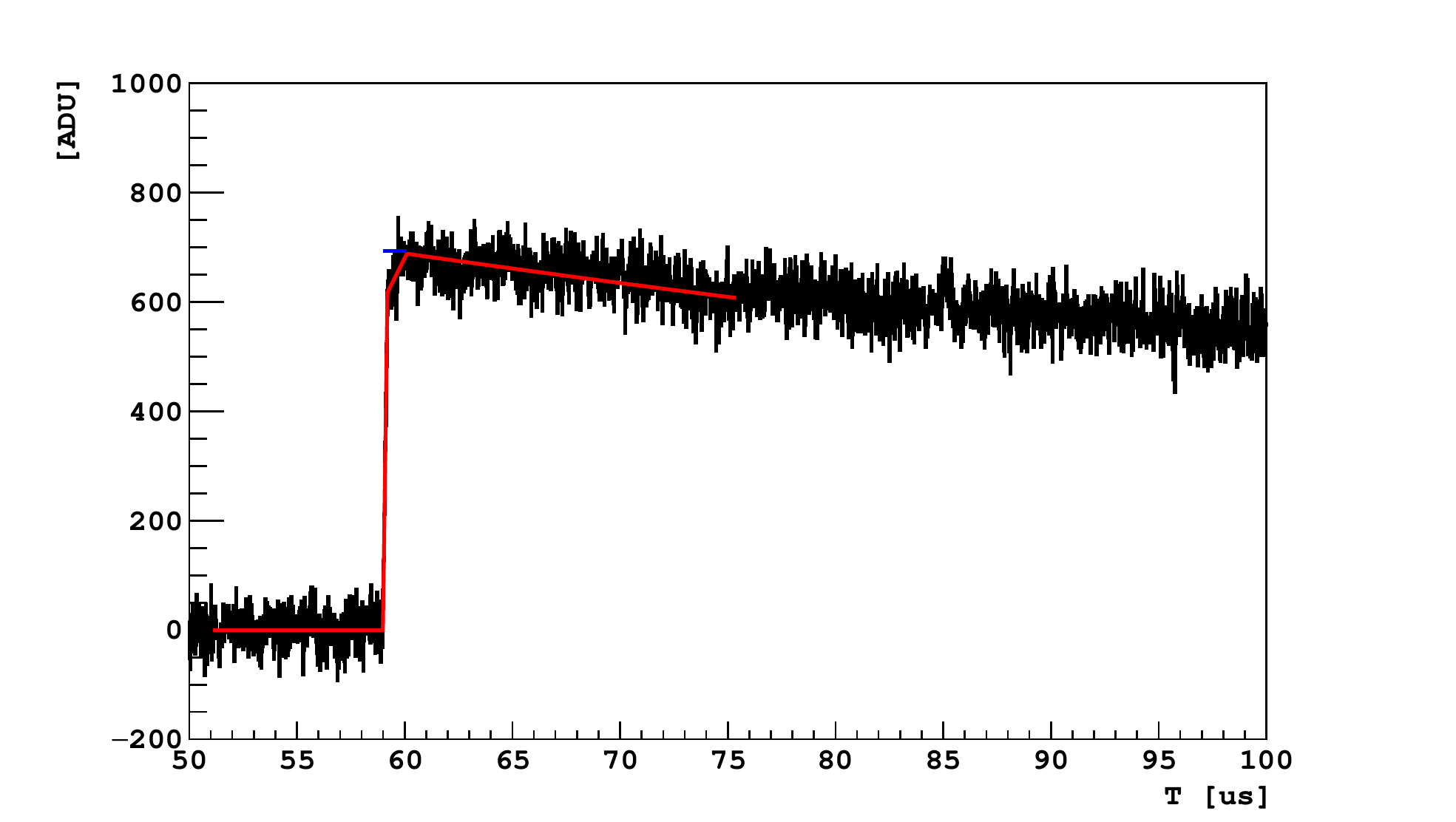}
        \vspace{-2em}
        \caption{}
        \end{subfigure}
        \begin{subfigure}[b]{0.75\linewidth}
        \includegraphics[width = \linewidth, trim={0 0 0 0.9cm}, clip]{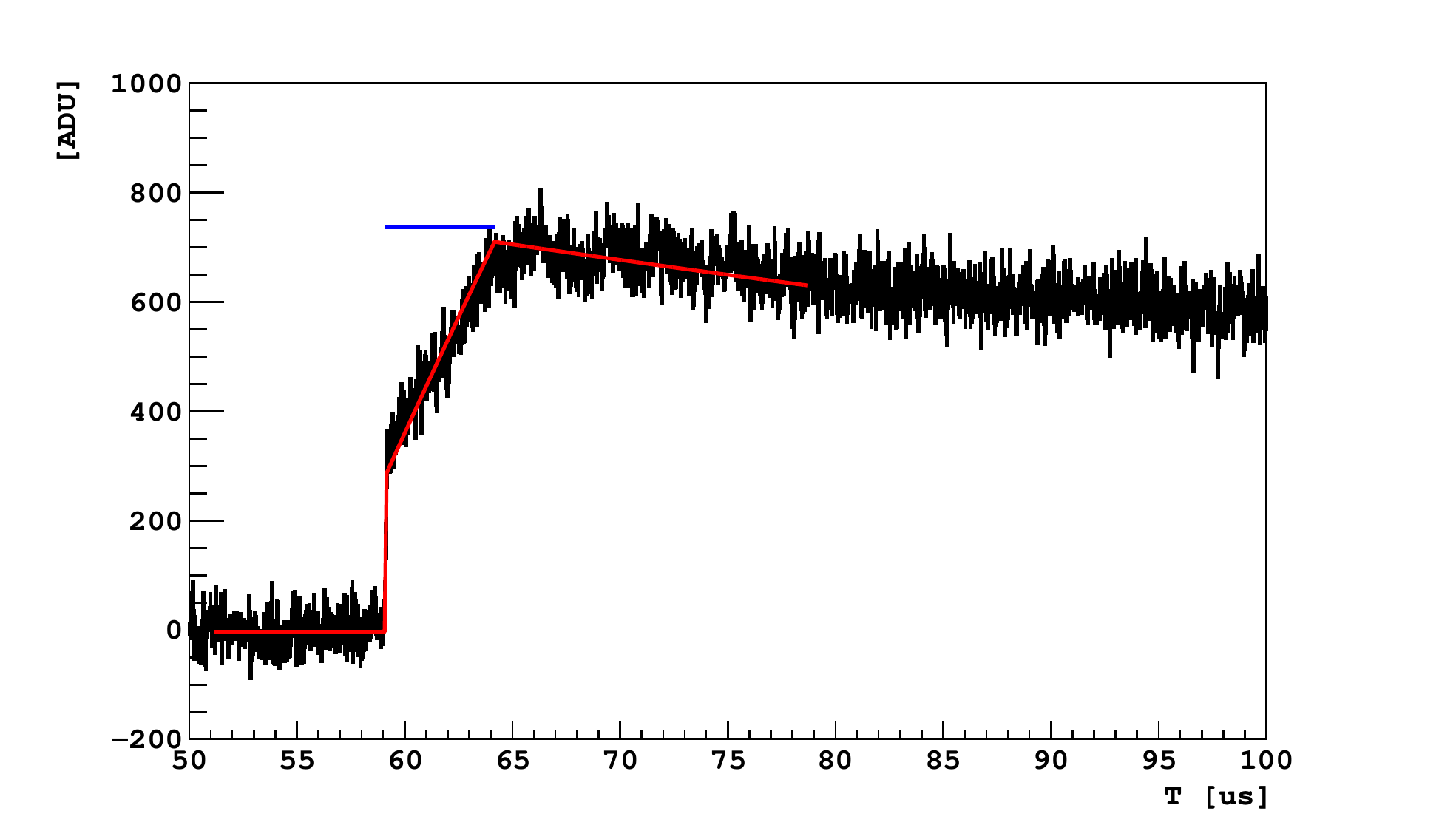}
        \vspace{-2em}
        \caption{}
        \end{subfigure}
        \begin{subfigure}[b]{0.75\linewidth}
        \includegraphics[width = \linewidth, trim={0 0 0 0.9cm}, clip]{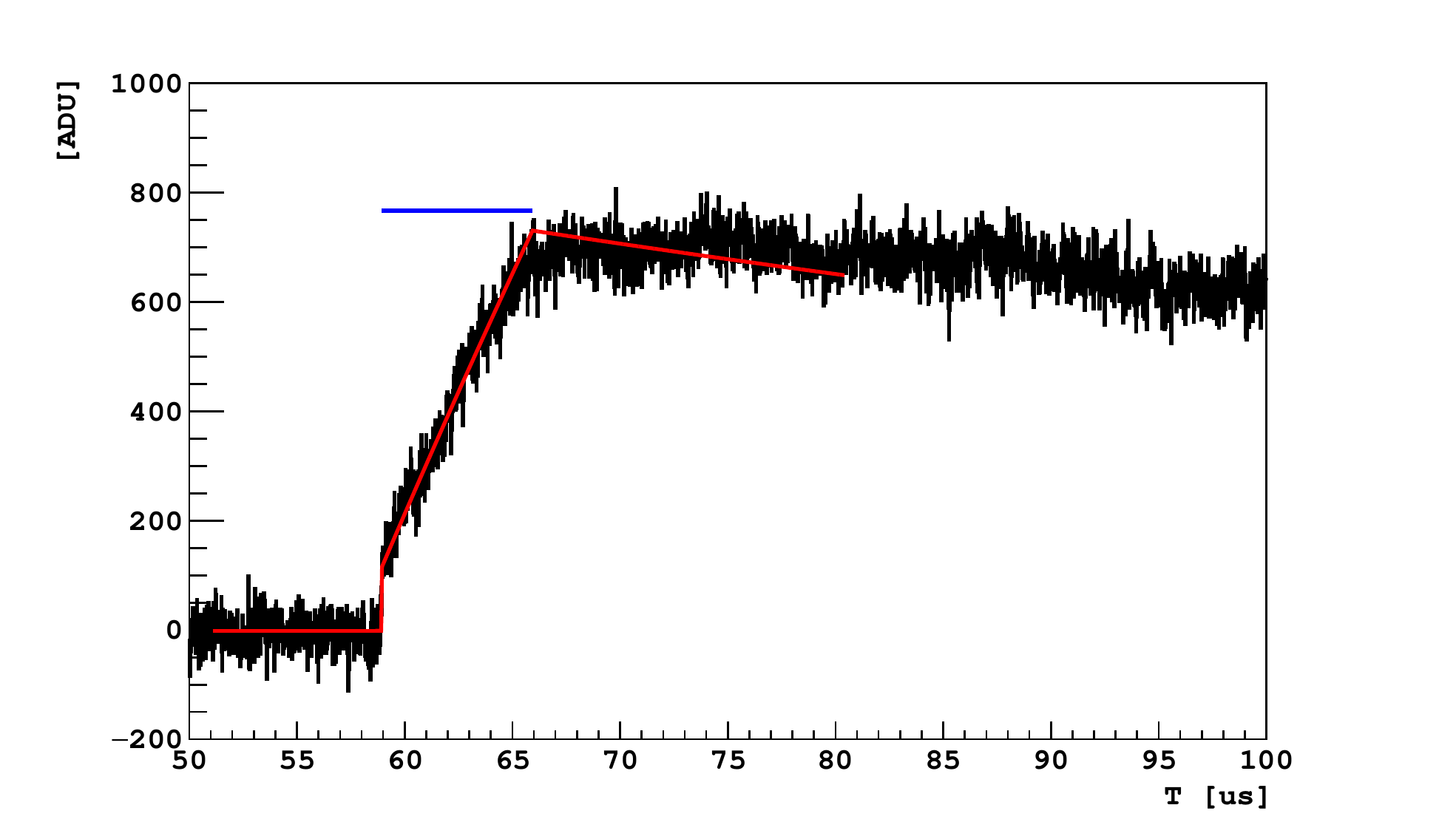}
        \vspace{-2em}
        \caption{}
        \end{subfigure}
        \caption{Recorded signal pulses (black) for ionization events at different DOI. The red curves show the fit result to the signal model in Equation~\ref{eq:SigModel}. The small discrepancy between the measured signal and the model is likely because of the assumption that the ionization event is point-like. The short horizontal blue line indicates the pulse amplitude for each event after correcting for signal decay. From top to bottom $\lambda$ increases from $\sim$0 (close to anode) to $\sim$1 (close to cathode). The pulses were acquired with $V_d=\SI{6}{\kilo\volt}$.}
        \label{fig:pulses}
    \end{figure}

The recorded pulses exhibit a \SI{111.2}{kHz} oscillation with steady phase and amplitude on the baseline from the switching frequency ripple of the HV power supply, whose amplitude became prominent above \SI{4}{kV} and kept increasing with increasing HV.
The oscillation was fit and removed from the recorded traces before further analysis.
Figure~\ref{fig:pulses} presents examples of the final signal pulses, with a baseline RMS noise of $\sim$270~$e^-$.

The ionization signal starts as soon as the e-h pairs start drifting in the aSe, which induce a current out of the cathode according to Ramo's theorem.
Drifting charges between two parallel planar electrodes generate a current
    \begin{equation}
    	I = q\mu \frac{V_d}{d^2}
    \end{equation}
where $q$ is the magnitude of the drifting charge, $V_d$ is the applied HV, $d$ is the separation between the electrodes, and $\mu$ is the charge-carrier mobility.
The output signal is then proportional to the integrated current convolved with the exponential decay constant $\tau$ because the output is capacitively coupled.
Thus, the signal may be expressed as 
    \begin{equation}
    	S(t) = GQ\frac{V_d}{d^2}\sum_{i=e,h}
        \begin{cases}
        0 & t<0 \\
        \mu_{i}\tau \big(1-e^{-t/\tau}\big) & 0<t<t_{i} \\
   \mu_{i}\tau \big(1-e^{-t_i/\tau}\big)e^{-\frac{t-t_{i}}{\tau}} & t>t_{i}
        \end{cases}
        \label{eq:SigModel}
    \end{equation}
where the sum is over the independent contribution to the signal from the drifting electrons ($e$) and holes ($h$).
$G$ is the calibration constant, $Q$ is the total free ionization charge, and $\mu_{e,h}$ is the mobility of the charge carriers.
We define $t=0$ as the time when the $\gamma$ ray interacts to generate free e-h pairs in the aSe, while $t_{e,h}$ is the time for electrons (holes) to drift to the anode (cathode).
Table~\ref{tab:aSepars} presents literature values for the charge carrier transport properties of aSe.
We do not include in Equation~\ref{eq:SigModel} the effect of the finite lifetime of the charge carriers in aSe ($\tau_{e,h}$) because it is significantly longer than the maximum carrier drift time.
We will revisit the effect of the finite electron lifetime on the measured signal amplitude in Section~\ref{sec:results}.

    \begin{table}[t]
        \centering
        \begin{tabular}{c|c c c}
        \hline
                     & Mobility
                     & Lifetime 
                     & Maximum drift time\\
                     & $[ \si{\um^2/V\us} ]$
                     & $[\si{\us}]$ 
                     & at $\SI{10}{V/\um}$ $[ \si{\us} ]$\\     
                     \hline
            Hole        & 13-16        &  10-40  &  1.3     \\
            Electron    & 0.6-0.8       &  40-50  &  30     \\
            \hline
        \end{tabular}
        \caption{Previously measured mobility and lifetime of charge carriers in aSe~\cite{Tabak1968}. The measured values, in particular the lifetime, depend on the sample quality and specific composition. The third column shows the expected maximum drift time in the aSe sensors for this experiment.}
        \label{tab:aSepars}
    \end{table}

The absorption of \SI{122}{\kilo\electronvolt} $\gamma$ rays is not exactly point-like because of the range of the photoelectrons and the emission of fluorescence X rays. However, our particle tracking simulations (see Section~\ref{sec:ionization}) show that $\sim$84\% of the ionization is generated over a range of \SI{20}{\micro\meter} in depth, which allows us to treat the events as point-like and define the depth of interaction (DOI) $\lambda$ as the average perpendicular distance of the ionization event from the anode as a fraction of $d$. The DOI is related to $t_{e,h}$ by
    \begin{align}
        t_e &= \lambda \frac{d^2}{\mu_{e}V_d} \\
        t_h &= (1-\lambda)\frac{d^2}{\mu_{h}V_d} 
    \end{align}

\section{Experimental results}
\label{sec:results}

Generally, $S(t)$ is the addition of the electron and hole signal components (Equation~\ref{eq:SigModel}).
However, when the interaction occurs very close to the anode (cathode), $t_e=0$ ($t_h=0$) and only the holes (electrons) contribute to the signal.
Such pulses can be approximated for $t\ll \tau$ by a linear function with the slope proportional to the carrier mobility.
Thus, we extract $\mu_h$ ($\mu_e$) from the pulses with the minimum (maximum) rise times in our data sample.

\begin{figure}
    \centering
    \begin{subfigure}[b]{0.8\linewidth}
    \includegraphics[width = \linewidth, trim={0 0 0 0.9cm}, clip]{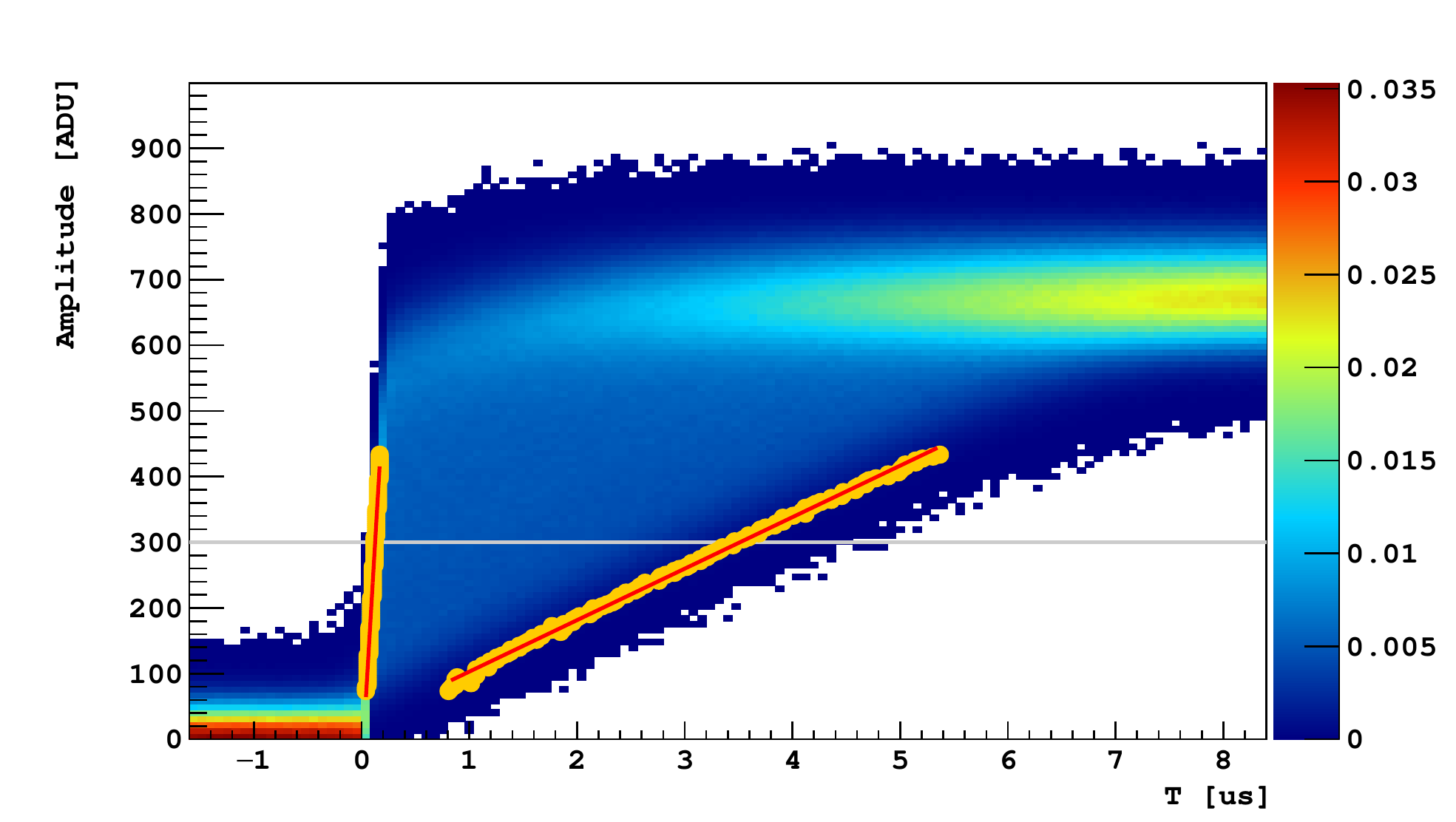}
    \vspace{-2em}
    \caption{}
    \end{subfigure}
    \begin{subfigure}[b]{0.8\linewidth}
    \includegraphics[width = \linewidth, trim={0 0 0 0.9cm}, clip]{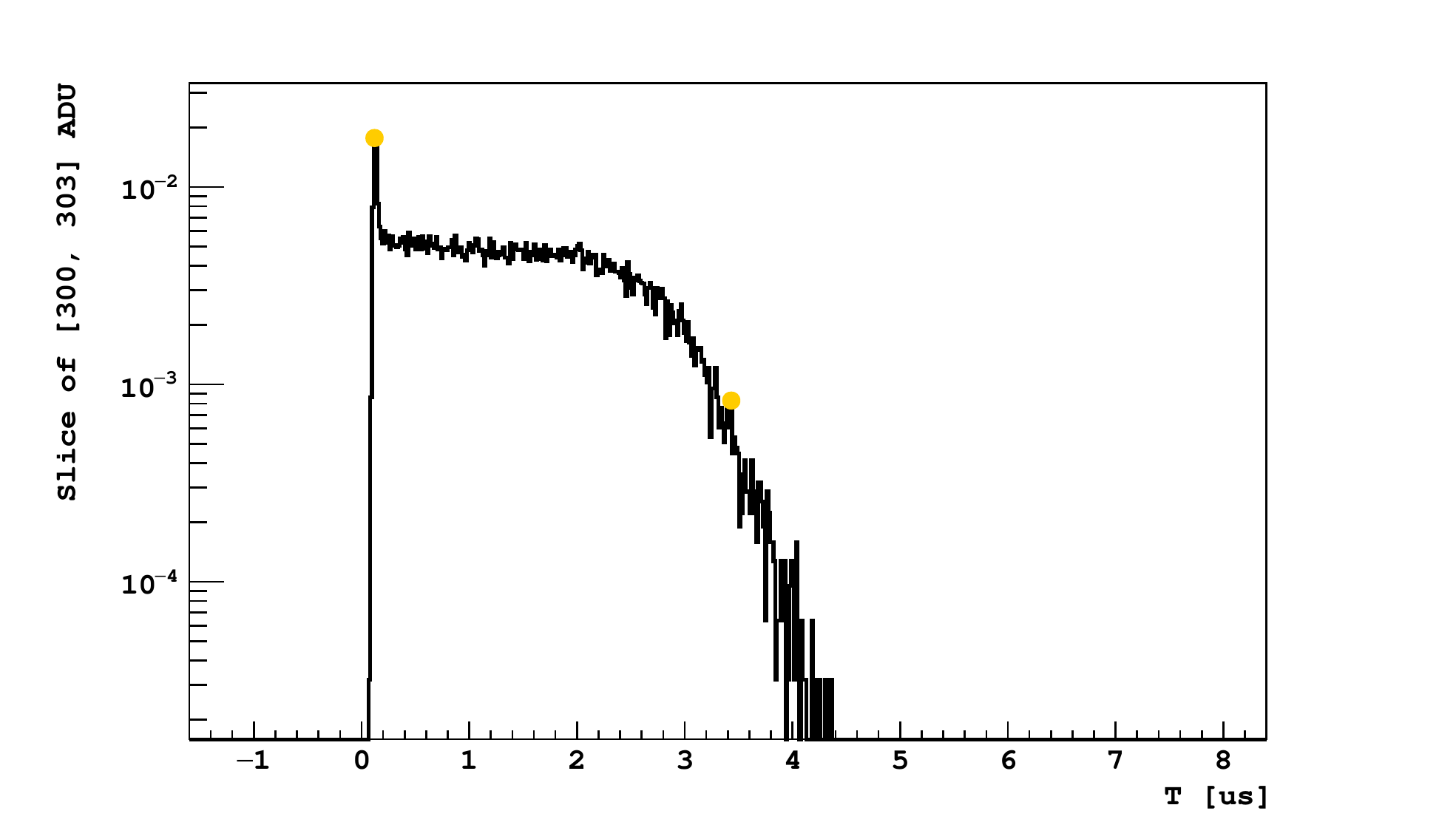}
    \vspace{-2em}
    \caption{}
    \end{subfigure}
    \caption{a)~Density plot of pulses with amplitudes within 10\% of the full absorption of \SI{122}{\kilo\electronvolt} $\gamma$ rays aligned in time for $E_d=$~\SI{30}{\volt\per\um}. Colors toward the red end of the spectrum correspond to higher densities. The average trace of pulses occurring very close to the anode (minimum rise time) and very close to the cathode (maximum rise time) are given by the circle markers, with the best-fit red lines to extract $\mu_{e,h}$ overlaid. b)~Distribution of pulse trace density as a function of time for the gray slice marked in panel (a) with pulse height in the range $[300,303]$ ADU. Circle markers denote the times that correspond to the average pulses with the minimum and maximum rise times.}
    \label{fig:AlignedWF30V}
\end{figure}

Pulses with height within $\sim$10\% of the full absorption energy of the primary \SI{122}{\kilo\electronvolt} $\gamma$ rays were selected and aligned in time such that the start of the pulse is at $t=0$.
The start of the pulse was defined as the intersection between the horizontal line defined by the mean of the baseline and the straight line obtained from the best linear fit to the first five consecutive samples that are above the baseline mean.
The baseline mean was subtracted before the pulses were added.
Figure~\ref{fig:AlignedWF30V}(a) shows a density plot of all the pulses acquired at $V_d=6$\,\si{\kilo\volt}, corresponding to a drift electric field $E_d =$~\SI{30}{\volt\per\um}.

\begin{figure}
    \centering
    \includegraphics[width = \linewidth]{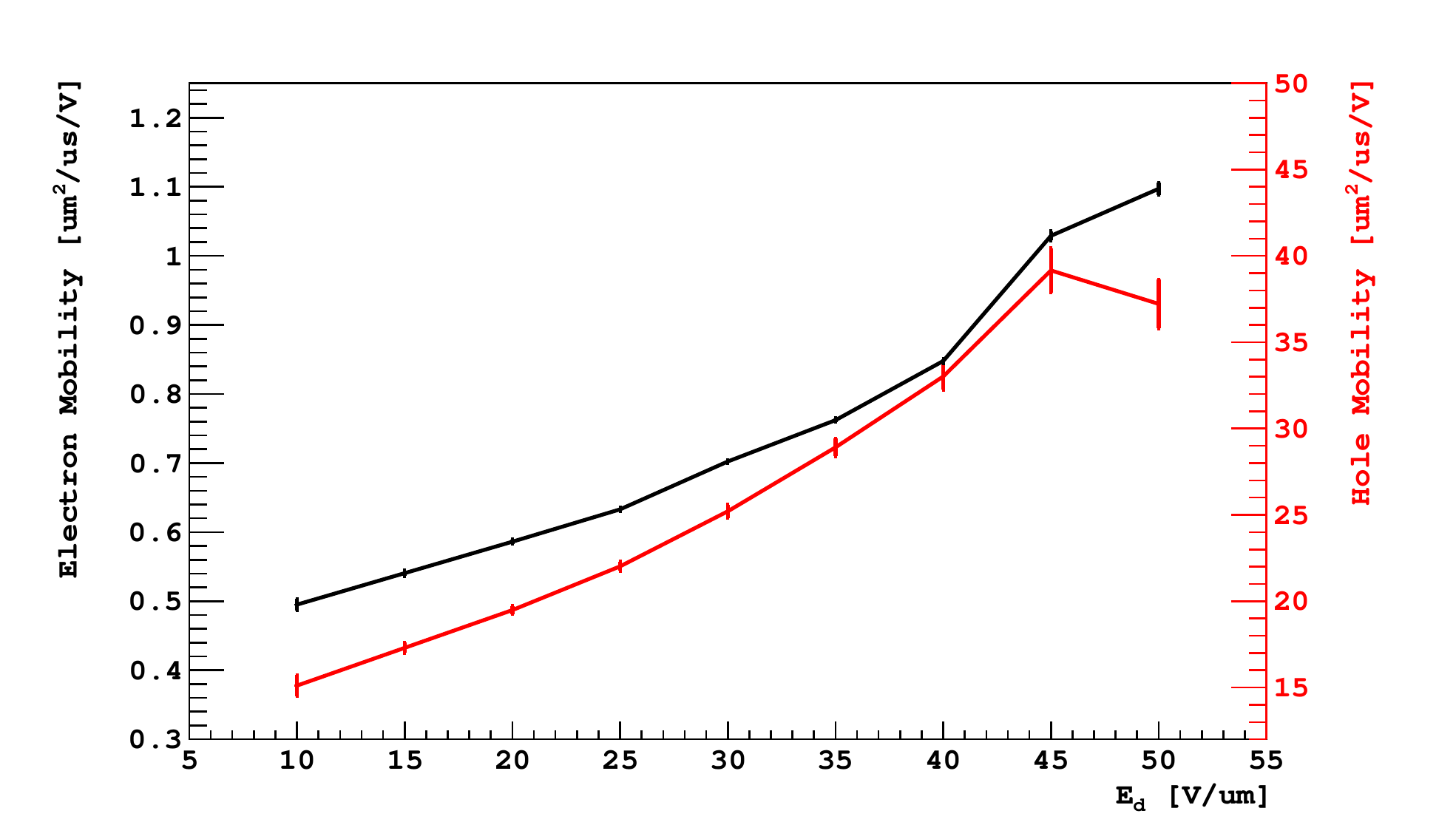}
    \caption{Mobility of electrons (black) and holes (red) in aSe as a fuction of $E_d$ obtained from the density plot of ionization pulses from \SI{122}{\kilo\electronvolt} $\gamma$ rays.}
    \label{fig:MueMuh}
\end{figure}

To construct the average time trace for pulses occurring very close to the anode and cathode, we plot the distribution as a function of time of the pulse trace density along horizontal slices (see Figure~\ref{fig:AlignedWF30V}(b)).
For each slice, we consider the time of the pulses with the minimum rise time to occur at the peak delineating the left edge of the distribution, while the maximum rise time is then estimated as the point on the right edge where the density becomes 10\% of the plateau level measured \SI{160}{\nano\second} after the peak.
The average pulses with minimum and maximum rise times are overlaid on the pulse-density plot as yellow circles in Figure~\ref{fig:AlignedWF30V}(a).
From a linear fit to the slope of the average pulses we obtain the charge carrier mobilities.
Figure~\ref{fig:MueMuh} shows measured values for $\mu_{e,h}$ as a function of $E_d$, in agreement with previous measurements~\cite{Juska1980}.
The value for $\mu_h$ plateaus above $E_d =$~\SI{45}{\volt\per\um} because the rise time of the pulses becomes comparable to the digitizer sampling time and, hence, the rise time cannot be measured any more precisely.

\begin{figure}
    \centering
    \includegraphics[width = \linewidth]{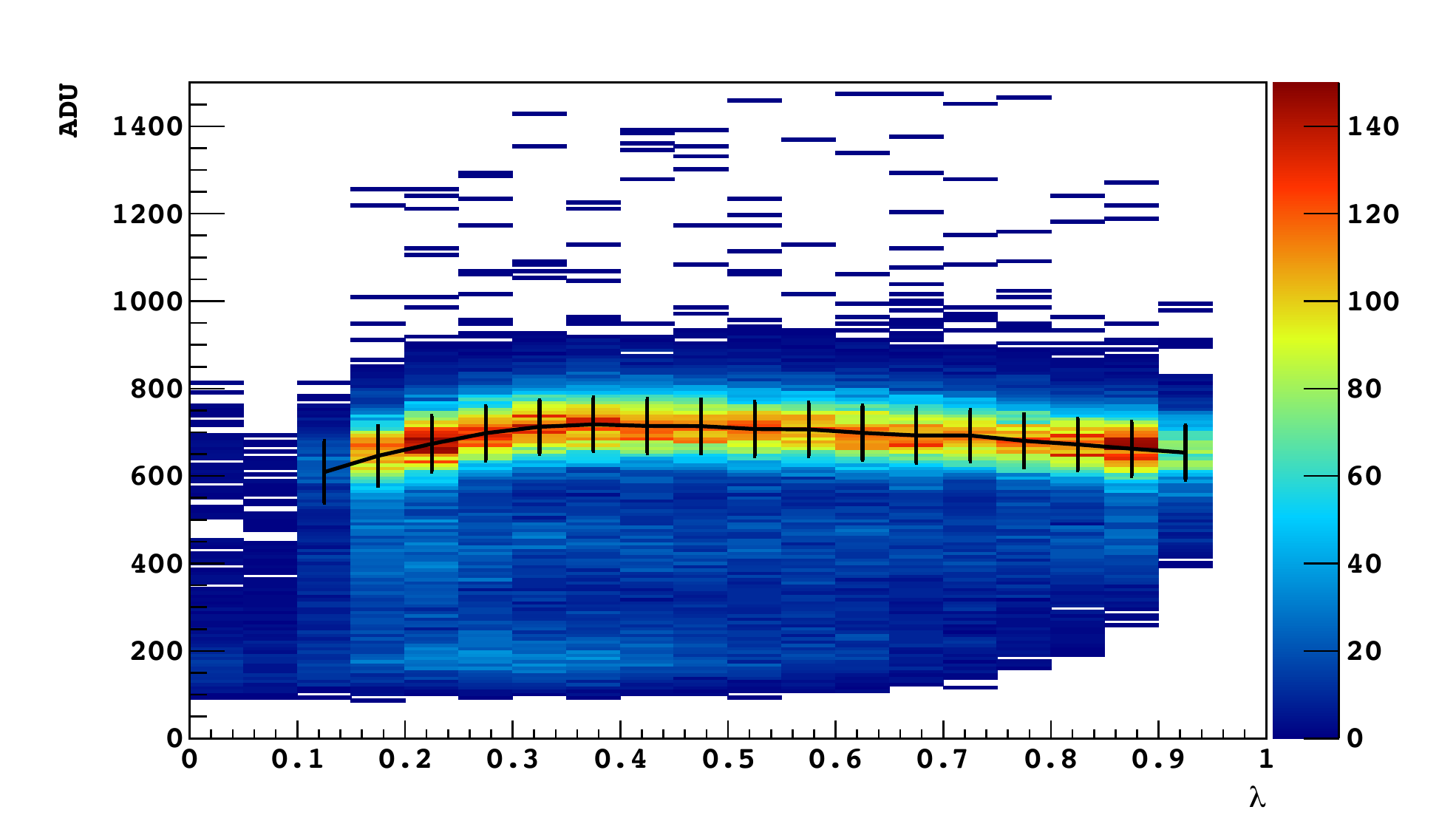}
    \caption{Two-dimensional histogram of $GQ$ versus $\lambda$ for \ce{$^{57}$Co} data acquired at a drift field $E_{d}=\SI{30}{V/\um}$. Color axis corresponds to the number of pulses in each bin. The black markers present the mean position of the full absorption peak as a function of $\lambda$, where the error bars represent the RMS width of the peak.}
    \label{fig:SpecHV30V}
\end{figure}

With the measured values of $\mu_{e,h}$, individual ionization pulses were fit with Equation~\ref{eq:SigModel} to extract the two remaining unknowns $Q$ and $\lambda$: the amplitude of the ionization signal and the DOI, respectively.
Figure~\ref{fig:SpecHV30V} shows a two-dimensional histogram of pulse height $GQ$ versus $\lambda$ for pulses acquired at $E_d =$~\SI{30}{\volt\per\um}.
The continuous horizontal band corresponds to the full absorption of the \SI{122}{\kilo\electronvolt} (\SI{85.5}{\percent} emission probability) and \SI{136}{\kilo\electronvolt} (\SI{10.7}{\percent} emission probability) $\gamma$ rays from \ce{$^{57}$Co}.
Since the aSe thickness is small compared to the \SI{0.5}{cm} attenuation length of the $\gamma$ rays, the DOI is evenly distributed across the layer.
A fraction of the $\gamma$ rays Compton scatter in the aSe, leading to a population of events with uniform DOI that extends to lower energies.
Some $\gamma$ rays Compton scatter in the collimator, leading to a population of lower energy events at small DOI.
Other lower energy photons emitted by the source are below the trigger threshold.
The events above the full absorption band may originate from higher-energy $\gamma$ rays from \ce{$^{57}$Co} with low emission probability or from environmental radiation, e.g., cosmic rays.
\begin{figure}
    \centering
    \includegraphics[width = \linewidth]{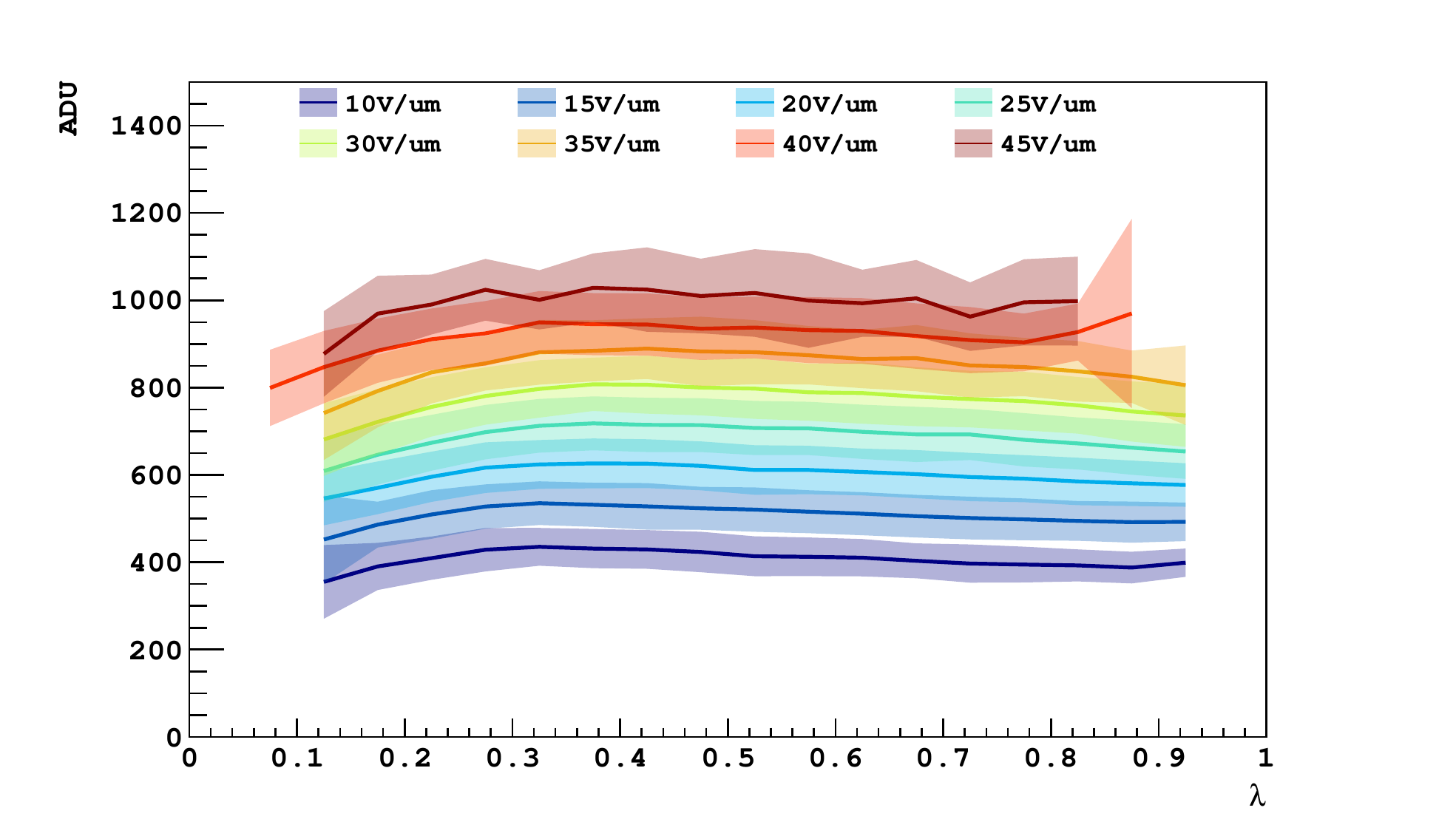}
    \caption{Full absorption peak as a function of $\lambda$ for different $E_d$. The color bands are centered at the mean peak position (solid lines) with the width of the band corresponding to the RMS width of the peak.}
    \label{fig:SpecSum}
\end{figure}

The markers in Figure~\ref{fig:SpecHV30V} show the result of a Gaussian fit to the spectrum in $GQ$ space for $\lambda$ bins of width 0.05.
Figure~\ref{fig:SpecSum} summarizes the result for different $E_{d}$, where the centroid of each band corresponds to the mean peak position and its width the standard deviation (RMS width) returned by the Gaussian fit.
The mean peak position increases with increasing $E_d$ because the charge-carrier recombination probability decreases with increasing electric field (see Section~\ref{sec:ionization}).
Although the peak position should be independent of DOI, we observe that the peak position for a given $E_d$ has a clear maximum at $\lambda\sim0.35$ with continuously decreasing values away from this point.
We expect a slight decrease in the peak position as a function of $\lambda$ because of the finite lifetime of the electrons $\tau_{e}$, which can become comparable to the maximum drift time (see Table~\ref{tab:aSepars}).
Once trapped, the electrons will not contribute to the amplitude of the pulse.
A constant $\tau_{e}$ should lead to a significant decrease in charge trapping with increasing $E_d$, since a lower drift time would lead to a smaller trapping probability.
However, the fractional decrease in the downward trend past the maximum observed in the data is similar for different $E_d$.
Thus, we were unable to remove the $\lambda$ dependence on the peak position by including $\tau_{e}$ in our model.
One possibility is that the charge trapping probability is not dependent on the charge-carrier drift time but on the distance that the charge carriers travel along the electric field direction.

Generally, almost any observed trend in the peak position can be explained by a corresponding trend in the drift electric field within the aSe.
For example, a smaller electric field near the anode could explain the lower peak position at small $\lambda$.
The hole blocking layer provides a smooth transition of field strength from zero at the anode to the maximum value in the bulk.
Because the hole blocking layer extends into the aSe, ionization events in this region experience greater recombination, as observed at lower $E_d$.
Likewise, fixed space charge within the aSe layer could explain the decreasing peak position toward the cathode.
In any case, we do not see any evidence of the aSe layer charging up with time, as we obtain consistent results from data acquired toward the beginning and the end of our longest (12-hour) acquisition run with $E_{d}=\SI{30}{\volt\per\um}$.

\begin{figure}
    \centering
    \includegraphics[width = \linewidth]{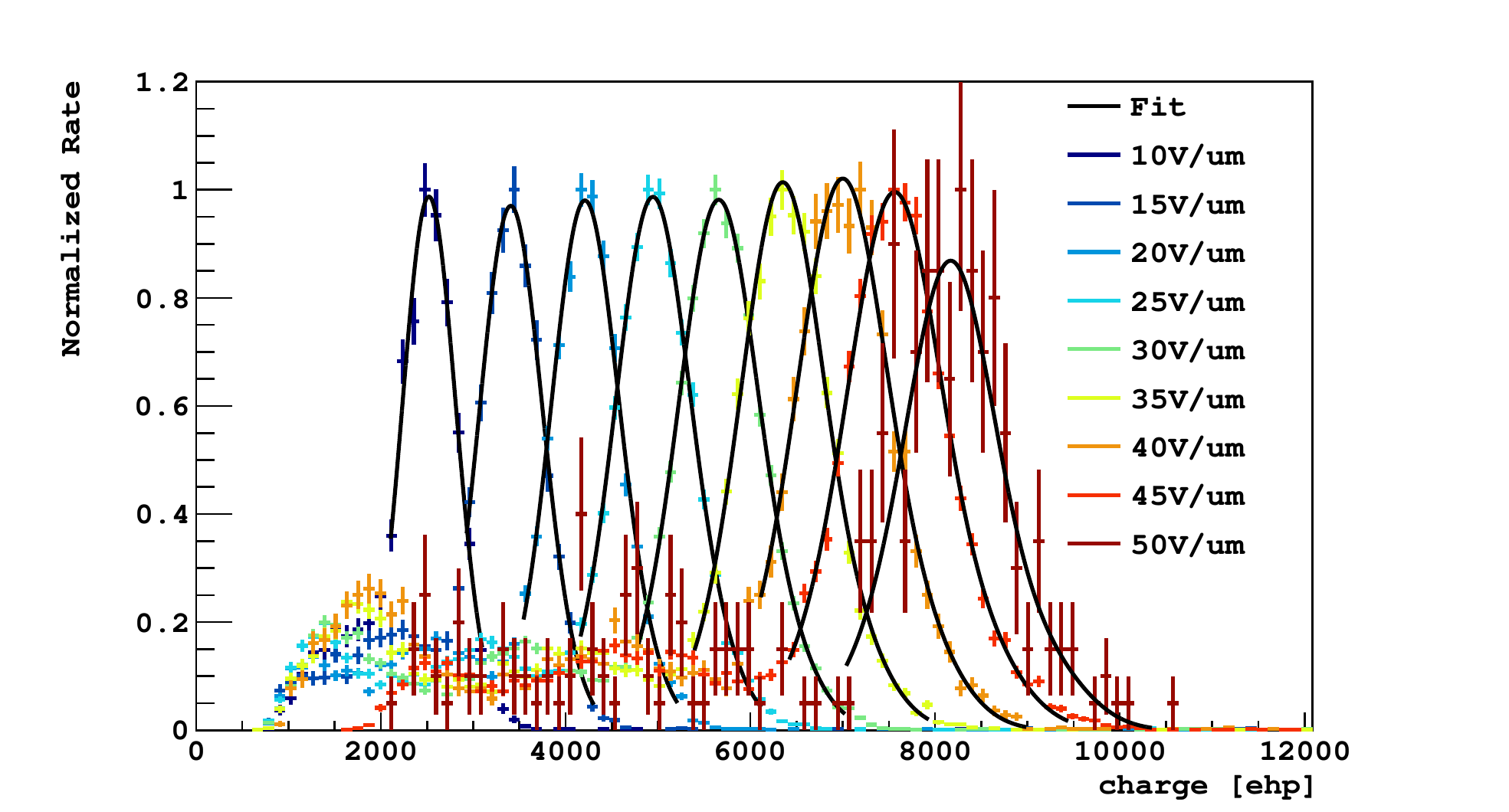}
    \caption{Measured spectra of events in the region $\lambda \in[0.25,0.5]$ under different $E_d$. The black lines are the best-fit result from the fit with Equation~\ref{eq:peak} to the full absorption peaks.}
    \label{fig:1DSpecSum}
\end{figure}

As we do not fully understand the DOI dependence of the full absorption peak, we do not attempt to correct for it and instead select the region  $\lambda \in[0.25,0.5]$ (approximately \SI{50}{\um} to \SI{100}{\um} from the anode) for further analysis.
We fit the peak observed in the $Q$ spectrum to a function with three Gaussian components:
\begin{equation}
    R(Q) = R_0 \frac{1}{\sqrt{2\pi\sigma^2}}\sum_{i=1}^{3} 
                b_{i} \exp{\left( -\frac{(Q-E_{i}/W)^2}{2 \sigma^2} \right)} 
    \label{eq:peak}
\end{equation}
where $W$ is the mean energy required to produce one free e-h pair in aSe, $E_1$ and $E_2$ account for the full absorption of the \SI{122}{\kilo\electronvolt} $\gamma$ rays and its corresponding K escape line at  \SI{110}{\kilo\electronvolt}, and $E_3$ accounts for the full absorption of \SI{136}{\kilo\electronvolt} $\gamma$ rays.
We use a common $\sigma$ for the RMS width of the three lines and constrain their relative intensities $b_i$ to the ratios $0.856\times0.9 : 0.856\times0.1 : 0.1068$.
The 10\% K X ray escape probability was obtained from a Geant4-based particle-tracking simulation that includes the detailed detector geometry (see Section~\ref{sec:ionization}).
The fit range was chosen as to avoid the lower energy tail of events from Compton scattering.
Figure~\ref{fig:1DSpecSum} presents the measured spectra at different $E_d$ with the overlaid best-fit line.

\section{Ionization response of amorphous selenium}
\label{sec:ionization}

To better understand the measured spectrum we developed a full detector simulation, including the microphysics of recombination in aSe.
The simulation is based on Geant4 \cite{Allison2016}, which tracks all particles and simulates all interactions starting with the emitted $\gamma$ rays from the source, followed by transport through the collimator and the device structure, the photoabsorption in the aSe layer, and the energy depositions by the ionizing electrons.
The simulation adopts the Penelope low energy physics that includes atomic relaxation processes \cite{Abdurashitov2016}.
For every energy deposition, we simulated the generation of free charge that drifts to the electrodes and gives rise to a signal pulse according to the model in Section~\ref{sec:signal}.
We introduced the baseline noise from the pre-trigger window in the data on the simulated pulses, and applied a digital low-pass filter with time constant $\tau$.
We performed the same signal extraction procedure on the simulated pulses as in the data to obtain the total pulse charge $Q$ for each event and generate simulated spectra from the \ce{^{57}Co} source.
Our detector simulation shows that our fitting procedure to the pulses successfully recovers $Q$ independent of $\lambda$ with an uncertainty from electronic noise and the signal extraction procedure of $\sigma_n=100$\,\si{e^-}.
This uncertainty contributes to the measured RMS width of the photabsorption lines as $\sigma^2=\sigma_n^2+\sigma_r^2$, where $\sigma_r$ is the contribution from fluctuations in charge generation in aSe.

For our charge-generation model, we set the ionizing electron step size in the aSe layer and the requirement to generate secondary electrons in Geant4 to their minimum values to best capture the fluctuations in the deposited-energy density along the track and to account for the large discrete energy depositions by low-energy secondary electrons. We take every step as a \SI{400}{\nano\meter}-long straight track segment and further divide it into subsegments of length $2r_0(E_K)$, where the ``spur size'' $r_0\sim$~1--5\,\si{\nano\meter} depends on the kinetic energy of the electron $E_K$ at the beginning of the step.
We take $r_0(E_K)$ to be proportional to the function in Figure~8 of Ref.~\cite{Fourkal2001} with scaling constant $a$ as a variable in our model.
The energy deposited by an ionizing electron in a step is calculated by Geant4 from a discrete number $N$ depositions of energy $\delta E$, where $N$ is determined by the mean free path of the electron as a function of $E_K$, and $\delta E$ follows a probability distribution proportional to $1/\delta E^2$~\cite{Lassila-Perini1995, Geant4Manual}.
We extracted these discrete energy depositions from Geant4 and distributed them randomly and uniformly along the step, adding the initial ionization charge from each energy deposit to the corresponding subsegment.
The initial ionization charge was calculated as $\delta E / w_0$, where the energy required to generate an e-h pair in aSe $w_0\sim$~4--7\,\si{\electronvolt}~\cite{Que1995, Klein1968} is a variable in our model.
For each subsegment, we follow the box recombination model from Ref.~\cite{Cataudella2017} to calculate the average free charge $\langle q \rangle$ that survives recombination assuming that the initial ionization charge $q_0$ is distributed uniformly within a cylindrical volume of length $2r_0$ and radius $r_0$:
\begin{equation}
    \langle q \rangle = q_0 \left( 1 + \frac{q_0 \alpha}{8\pi r_0^2 E_d \mu(E_d)}  \right)^{-1}
    \label{eq:recomb}
\end{equation}
For the recombination coefficient $\alpha$, we adopt the modified expression proposed for aSe in Ref.~\cite{Bubon:2016hc}:
\begin{equation}
    \alpha = \left( \frac{\epsilon}{e \mu(E_d)} + \frac{1}{\alpha_0} \right)^{-1}
    \label{eq:alpha}
\end{equation}
with $\alpha_0=$~\SI{5E-8}{\cm\cubed\per\second} and the permittivity of aSe $\epsilon=6.3\epsilon_0$~\cite{Lachaine2000,Bubon:2016hc}, where $\epsilon_0$ is the vacuum permittivity.
In Equations~\ref{eq:recomb} and \ref{eq:alpha}, $\mu(E_d)$ corresponds to the sum of the mobilities of the charge carriers, which in aSe can be approximated as $\mu(E_d)\sim\mu_h(E_d)$ measured in Section~\ref{sec:results}.
For every track subsegment, we assume that charge $q$ is generated, drawn from a Poisson distribution with mean $\langle q \rangle$.
Following the complete detector simulation, we obtained predicted spectra for different values of $E_d$, and recombination model parameters $a$ and $w_0$.

\begin{figure}
    \centering
    \includegraphics[width = \linewidth]{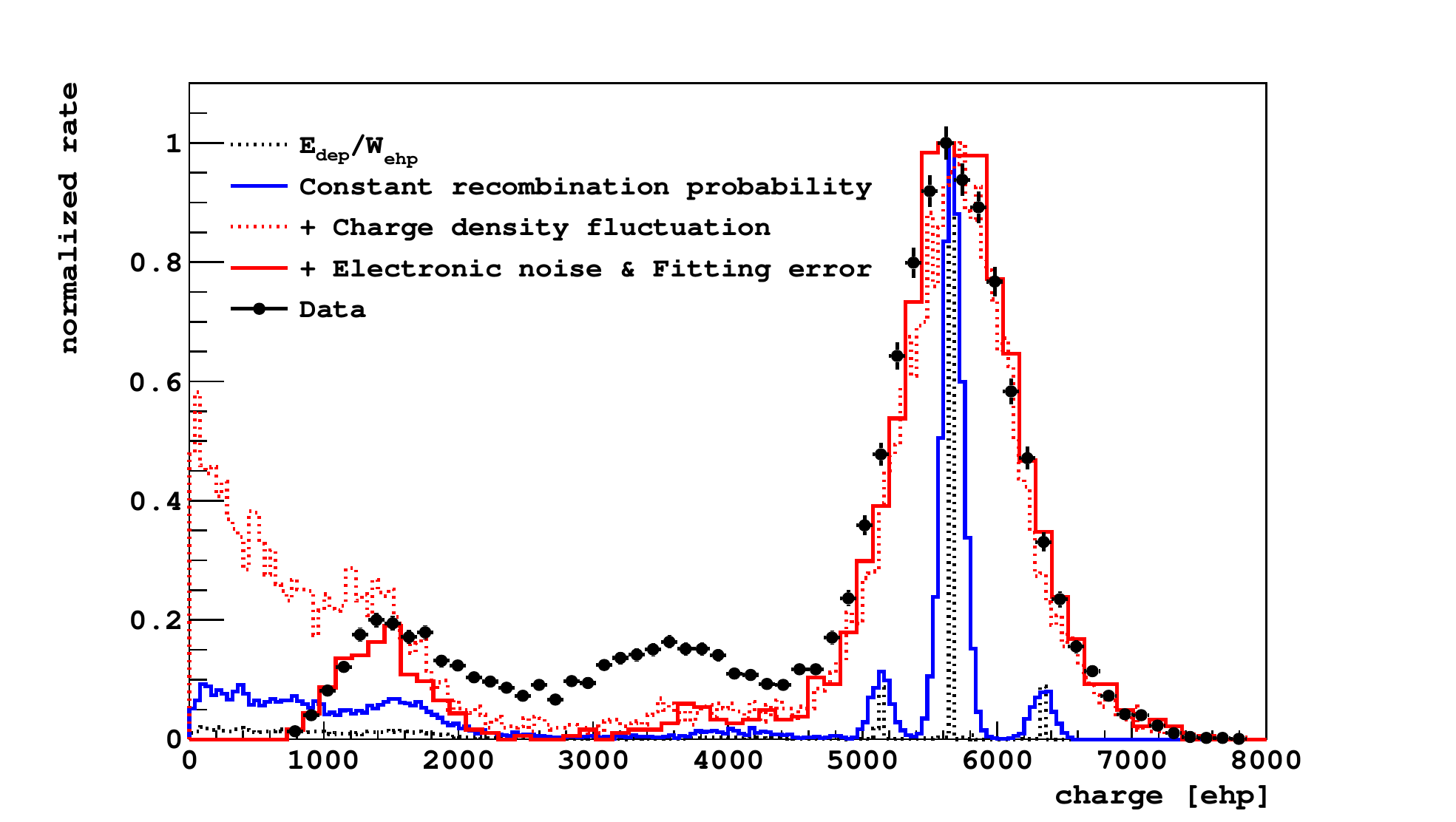}
    \caption{Predicted energy spectrum by our detector simulation with $a=1.33$ and $w_0=5.0$\,\si{\electronvolt} for $E_d=$~\SI{30}{\volt\per\micro\meter}. The data (black markers) are well reproduced by the simulation (red line). We also present the deposited-energy spectrum in the aSe (dashed black line) and the simulation result before the contribution from electronic noise and signal extraction (doted red line). The predicted spectrum if the recombination probability were constant and independent for every generated e-h pair is shown by the blue line.}
    \label{fig:FluctComp}
\end{figure}
Figure~\ref{fig:FluctComp} presents an example of the predicted energy spectrum from our simulation.
The dominant contribution to the width of the full absorption peak comes from the dependence of the recombination probability on the deposited-energy density along the ionizing-electron tracks.
The figure also shows the predicted spectrum if the recombination probability were to be constant and uncorrelated for every generated e-h pair, exhibiting a much better energy resolution capable of resolving the three separate spectral lines.
The amplitude of the low-energy tail on the left-hand side of the peak is larger in the data than in the simulation possibly because the alignment of the collimator is more accurate in the simulation than in the real experiment.
The cutoff in the energy spectrum for $Q<1000$ is caused by the trigger threshold, which is well reproduced by the electronics simulation.

\begin{figure}
    \centering
    \begin{subfigure}[b]{0.8\linewidth}
        \includegraphics[width = \linewidth, trim={0 0 0 0.9cm}, clip]{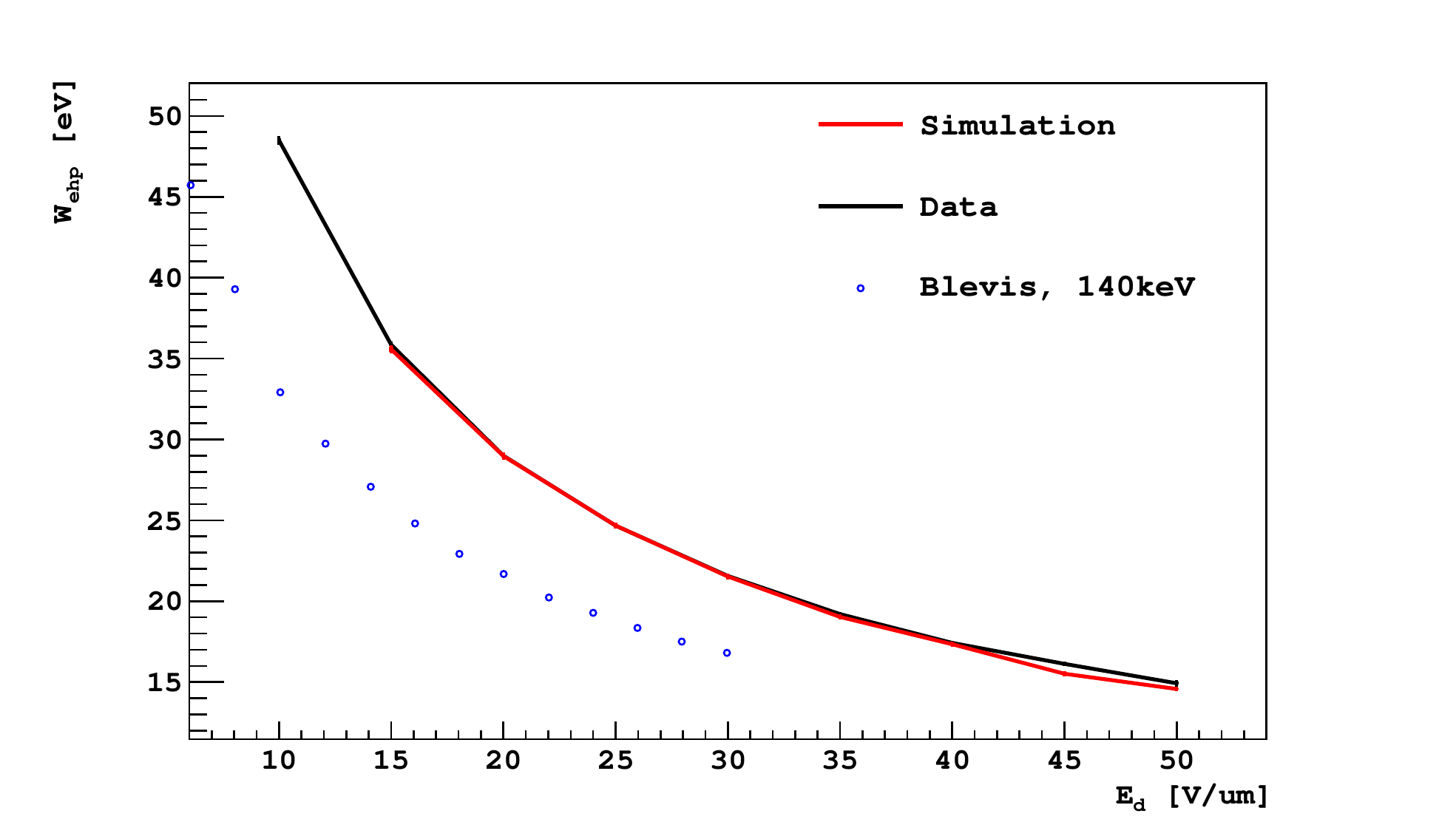}
        \vspace{-2em}
    \caption{}
    \end{subfigure}
    \begin{subfigure}[b]{0.8\linewidth}
        \includegraphics[width = \linewidth, trim={0 0 0 0.9cm}, clip]{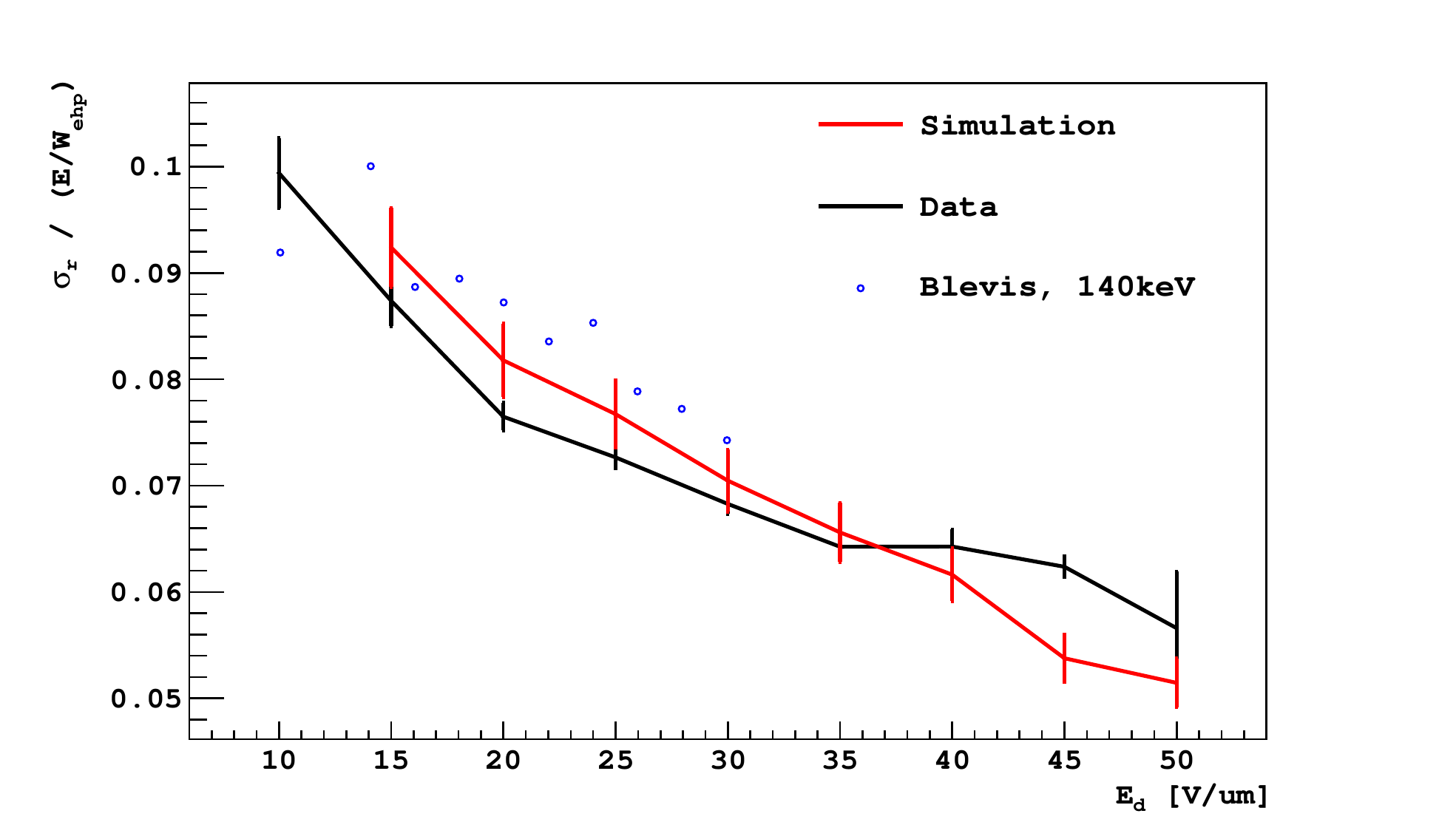}
        \vspace{-2em}
    \caption{}
    \end{subfigure}
    \caption{a)~Average energy to produce a free e-h pair in aSe $W$ as a function of drift electric field $E_d$ measured from the fit to the photoabsorption peak. The error bars represent statistical uncertainty only. The systematic uncertainty from the gain of the electronics chain is $\sim$\SI{5}{\percent}. b)~Fractional resolution of the photoabsorption line as a function of $E_d$. Black markers correspond to our measurement performed with \SI{122}{\kilo\electronvolt} $\gamma$ rays, while blue markers present past measurements performed with \SI{140}{\kilo\electronvolt} X rays. The red markers are the results from the microphysics simulation described in the text.}
    \label{fig:SimAndData}
\end{figure}
Figure~\ref{fig:SimAndData} summarizes the measured values of $W$ and $\sigma_r/E_1$, the fractional energy resolution at \SI{122}{\kilo\electronvolt}, as a function of $E_d$.
These two parameters are the most relevant to understand the energy response of aSe.
The value of $W$ decreases with increasing $E_d$ because of the decrease in the fraction of primary e-h pairs that recombine.
This trend was observed in past measurements~\cite{Bubon:2016hc}, although the reported absolute values of $W$ at a particular $E_d$ appear to vary from sample to sample depending on the specific composition of the aSe layer~\cite{Kasap2003}.
The observed energy resolution is comparable to past measurements with \SI{140}{\kilo\electronvolt} X rays~\cite{Blevis:1999ht}.
From our simulations, we obtain predicted curves for $W(E_d)$ as a function of $a$ and $w_0$.
From a fit to the data shown by the red line in Figure~\ref{fig:SimAndData}(a), we obtained best-fit values $a=1.33 \pm 0.04$ and $w_0=5.0 \pm 0.3$\,\si{\electronvolt}.
The corresponding prediction for $\sigma_r/E_1$ is shown by the red line in Figure~\ref{fig:SimAndData}(b), in satisfactory agreement with data.

\section{Implications for neutrinoless $\beta \beta$ decay search}
\label{sec:bb}

The recombination model presented in Section~\ref{sec:ionization} provides a framework to estimate the energy resolution of aSe to events from the neutrinoless \bb\ decay of \ce{$^{82}$Se}.
We extended the size of the aSe target layer in our Geant4-based simulation to fully contain pairs of electrons from neutrinoless \bb\ decay.
Starting from the initial kinetic energy distributions provided by the Decay0 program~\cite{Ponkratenko:2000im}, we simulated the ionizing electron tracks and the corresponding charge generation and recombination processes.
From the distribution of the total free charge $Q$ produced in each event, we obtain an energy resolution at $Q_{\beta\beta}=3.0$\,\si{\mega\electronvolt} of 2.8\% for $E_d=$~\SI{30}{\volt\per\micro\meter}.
This result is significantly larger than the 0.34\% estimated in Ref.~\cite{Chavarria:2016vw}.
Our simulation predicts that increasing the drift electric field could further improve the energy resolution to 2.0\% at $E_d=$~\SI{50}{\volt\per\micro\meter}.
Further, because fluctuations in ionization density are correlated with straggling of the electrons, the average charge yield for an electron track is anti-correlated with its length.
Hence, if the track length could be accurately measured (as in the detector proposed in Ref.~\cite{Chavarria:2016vw}), with a simple first-order linear correction the resolution could be improved to 1.1\%.

Note that our recombination model has only been validated with $\sim$\SI{100}{\kilo\electronvolt} electrons from the photoabsorption of $\gamma$ rays from a \ce{^{57}Co} source, while the electrons emitted by \bb\ decay are an order of magnitude more energetic.
In particular, the deposited-energy density is significantly lower along the tracks of higher-energy electrons, an unexplored regime where we expect a smaller charge recombination probability.
The experimental technique used for this measurement cannot be extended to electrons of much higher energy because above $\sim$\SI{200}{\kilo\electronvolt} $\gamma$ rays primarily Compton scatter, generating multiple low-energy electrons, and the few high-energy electrons from photoelectric absorption escape the aSe target because of their long range.
We are fabricating larger hybrid aSe/CMOS imaging sensors \SI{36}{\mm\squared} in area with \SI{500}{\micro\meter}-thick aSe layers that can fully contain  higher-energy electrons.
These devices will allow us to measure the charge yield as a function of deposited-energy density for a better understanding of the ionization response of aSe.
Furthermore, we expect to improve the energy resolution by correcting for the difference in charge yield along the imaged electron tracks.
Beyond simple analytical corrections, this problem is particularly well suited for machine learning approaches~\cite{Ai_2018, Liu:2020pzv} given the wealth of information encoded in the high-resolution electron tracks.
Ultimately, we aim to implement a detailed ionization response model and energy reconstruction procedure in a realistic simulation of the detector proposed in Ref.~\cite{Chavarria:2016vw} to accurately evaluate its prospects in the search for neutrinoless \bb\ decay.

\acknowledgments{This material is based upon work supported by the U.S. Department of Energy, Office of Science, Office of Nuclear Physics Fundamental Symmetries program under Award Number DE-SC-0020439.}

\bibliographystyle{jhep}
\bibliography{myrefs}

\providecommand{\href}[2]{#2}\begingroup\raggedright\begin{thebibliography}{10}

\bibitem{Chavarria:2016vw}
A.~E. Chavarria, C.~Galbiati, X.~Li and J.~A. Rowlands, \emph{{A
  high-resolution CMOS imaging detector for the search of neutrinoless double
  $\beta$ decay in $^{82}$Se}},
  \href{http://dx.doi.org/10.1088/1748-0221/12/03/P03022}{\emph{J. Instrum.}
  {\bf 12} (2017) P03022}.

\bibitem{DellOro:2016gf}
S.~Dell{\textquoteright}Oro, S.~Marcocci, M.~Viel and F.~Vissani,
  \emph{{Neutrinoless Double Beta Decay: 2015 Review}},
  \href{http://dx.doi.org/10.1155/2016/2162659}{\emph{Adv. High Energy Phys.}
  {\bf 2016} (2016) 2162659}.

\bibitem{Tabak1968}
M.~D. Tabak and P.~J. Warter, \emph{{Field-controlled photogeneration and
  free-carrier transport in amorphous selenium films}},
  \href{http://dx.doi.org/10.1103/PhysRev.173.899}{\emph{Phys. Rev.} {\bf 173}
  (1968) 899}.

\bibitem{Juska1980}
G.~Ju{\v{s}}ka and K.~Arlauskas, \emph{{Impact ionization and mobilities of
  charge carriers at high electric fields in amorphous selenium}},
  \href{http://dx.doi.org/10.1002/pssa.2210590151}{\emph{Phys. Status Solidi A}
  {\bf 59} (1980) 389}.

\bibitem{Allison2016}
J.~Allison, K.~Amako, J.~Apostolakis, P.~Arce, M.~Asai, T.~Aso et~al.,
  \emph{{Recent developments in GEANT4}},
  \href{http://dx.doi.org/10.1016/j.nima.2016.06.125}{\emph{Nucl. Instrum.
  Methods Phys. Res. A} {\bf 835} (2016) 186}.

\bibitem{Abdurashitov2016}
D.~N. Abdurashitov, Y.~M. Malyshkin, V.~L. Matushko and B.~Suerfu,
  \emph{{Response of a proportional counter to $^{37}$Ar and $^{71}$Ge:
  Measured spectra versus Geant4 simulation}},
  \href{http://dx.doi.org/10.1016/j.nimb.2016.02.029}{\emph{Nucl. Instrum.
  Methods Phys. Res. B} {\bf 373} (2016) 5}.

\bibitem{Fourkal2001}
E.~Fourkal, M.~Lachaine and B.~G. Fallone, \emph{{Signal formation in
  amorphous-Se-based x-ray detectors}},
  \href{http://dx.doi.org/10.1103/PhysRevB.63.195204}{\emph{Phys. Rev. B} {\bf
  63} (2001) 195204}.

\bibitem{Lassila-Perini1995}
K.~Lassila-Perini and L.~Urb{\'{a}}n, \emph{{Energy loss in thin layers in
  GEANT}}, \href{http://dx.doi.org/10.1016/0168-9002(95)00344-4}{\emph{Nucl.
  Instrum. Methods Phys. Res. A} {\bf 362} (1995) 416}.

\bibitem{Geant4Manual}
{\scshape {Geant4}} collaboration, ``{Energy Loss Fluctuations \textemdash\
  Physics Reference Manual 10.4 documentation}.''
  \url{http://geant4-userdoc.web.cern.ch/geant4-userdoc/UsersGuides/PhysicsReferenceManual/BackupVersions/V10.4/html/electromagnetic/energy\_loss/fluctuations.html\#em-straggling-bichsel}.
\newblock Accessed: 2020-10-10.

\bibitem{Que1995}
W.~Que and J.~A. Rowlands, \emph{{X-ray photogeneration in amorphous selenium:
  Geminate versus columnar recombination}},
  \href{http://dx.doi.org/10.1103/PhysRevB.51.10500}{\emph{Phys. Rev. B} {\bf
  51} (1995) 10500}.

\bibitem{Klein1968}
C.~A. Klein, \emph{{Bandgap dependence and related features of radiation
  ionization energies in semiconductors}},
  \href{http://dx.doi.org/10.1063/1.1656484}{\emph{J. Appl. Phys.} {\bf 39}
  (1968) 2029}.

\bibitem{Cataudella2017}
V.~Cataudella, A.~{De Candia}, G.~D. Filippis, S.~Catalanotti, M.~Cadeddu,
  M.~Lissia et~al., \emph{{Directional modulation of electron-ion pairs
  recombination in liquid argon}},
  \href{http://dx.doi.org/10.1088/1748-0221/12/12/P12002}{\emph{J. Instrum.}
  {\bf 12} (2017) P12002}.

\bibitem{Bubon:2016hc}
O.~Bubon, K.~Jandieri, S.~D. Baranovskii, S.~O. Kasap and A.~Reznik,
  \emph{{Columnar recombination for X-ray generated electron-holes in amorphous
  selenium and its significance in a-Se x-ray detectors}},
  \href{http://dx.doi.org/10.1063/1.4944880}{\emph{J. Appl. Phys.} {\bf 119}
  (2016) 124511}.

\bibitem{Lachaine2000}
M.~Lachaine and B.~G. Fallone, \emph{{Monte Carlo simulations of x-ray induced
  recombination in amorphous selenium}},
  \href{http://dx.doi.org/10.1088/0022-3727/33/11/323}{\emph{J. Phys. D: Appl.
  Phys.} {\bf 33} (2000) 1417}.

\bibitem{Kasap2003}
S.~O. Kasap, K.~V. Koughia, B.~Fogal, G.~Belev and R.~E. Johanson, \emph{{The
  influence of deposition conditions and alloying on the electronic properties
  of amorphous selenium}},
  \href{http://dx.doi.org/10.1134/1.1592851}{\emph{Semiconductors} {\bf 37}
  (2003) 789}.

\bibitem{Blevis:1999ht}
I.~M. Blevis, D.~C. Hunt and J.~A. Rowlands, \emph{{Measurement of x-ray
  photogeneration in amorphous selenium}},
  \href{http://dx.doi.org/10.1063/1.370615}{\emph{J. Appl. Phys.} {\bf 85}
  (1999) 7958}.

\bibitem{Ponkratenko:2000im}
O.~A. Ponkratenko, V.~I. Tretyak and Y.~G. Zdesenko, \emph{{Event generator
  DECAY4 for simulating double-beta processes and decays of radioactive
  nuclei}}, \href{http://dx.doi.org/10.1134/1.855784}{\emph{Phys. At. Nucl.}
  {\bf 63} (2000) 1282}.

\bibitem{Ai_2018}
P.~Ai, D.~Wang, G.~Huang and X.~Sun, \emph{{Three-dimensional convolutional
  neural networks for neutrinoless double-beta decay signal/background
  discrimination in high-pressure gaseous Time Projection Chamber}},
  \href{http://dx.doi.org/10.1088/1748-0221/13/08/p08015}{\emph{J. Instrum.}
  {\bf 13} (2018) P08015}.

\bibitem{Liu:2020pzv}
J.~Liu, J.~Ott, J.~Collado, B.~Jargowsky, W.~Wu, J.~Bian et~al.,
  \emph{{Deep-Learning-Based Kinematic Reconstruction for DUNE}},
  \href{https://arxiv.org/abs/2012.06181}{{\tt 2012.06181}}.

\end{thebibliography}\endgroup
\end{document}